\documentclass{emulateapj}
\usepackage[]{times, graphicx}
\newcommand{\soho}{{\em SOHO{}}}
\newcommand{\hinode}{{\em Hinode}}

\newcommand{\pref}{\protect\ref}
\begin{document}

\shorttitle{The Dynamic Transition Region}
\shortauthors{McIntosh \& De~Pontieu}
%\title{The Transition Region: A Result of Episodic Heating and Relentless Mass Transport Between the Chromosphere and Corona}
\title{Ubiquitous High Speed Transition Region and Coronal Upflows in the Quiet Sun}
%\title{The Solar Transition Region: A Mass Conveyor Between the Chromosphere and Corona} 

\author{Scott W. McIntosh\altaffilmark{1,3}, Bart De Pontieu\altaffilmark{2,3}} 
\altaffiltext{1}{High Altitude Observatory, National Center for Atmospheric Research, P.O. Box 3000, Boulder, CO 80307}
\altaffiltext{2}{Lockheed Martin Solar and Astrophysics Lab, 3251 Hanover St., Org. ADBS, Bldg. 252, Palo Alto, CA  94304}
\altaffiltext{3}{These authors contributed equally to the production of this manuscript.}
\email{mscott@ucar.edu, bdp@lmsal.com}

\begin{abstract}
We study the line profiles of a range of transition region (TR) emission lines observed in typical quiet Sun regions. 
In magnetic network regions, the \ion{Si}{4}~1402 \AA{}, \ion{C}{4}~1548 \AA{}, \ion{N}{5}~1238 \AA{}, \ion{O}{6}~1031\AA{}, and \ion{Ne}{8}~770 \AA{} spectral lines show significant asymmetry in the blue wing of the emission line profiles. We interpret these high-velocity upflows in the lower and upper TR as the quiet Sun equivalent of the recently discovered upflows in the low corona above plage regions \citep{Hara2008}. The latter have been shown to be directly associated with high-velocity chromospheric spicules that are (partially) heated to coronal temperatures and play a significant role in supplying the active region corona with hot plasma \citep{DePontieu2009}. We show that a similar process likely dominates the quiet Sun network. We provide a new interpretation of the observed quiet Sun TR emission in terms of the relentless mass transport between the chromosphere and corona - a mixture of emission from dynamic episodic heating and mass injection into the corona as well as that from the previously filled, slowly cooling, coronal plasma. Analysis of the observed upflow component shows that it carries enough hot plasma to play a significant role in the energy and mass balance of the quiet corona. We determine the temperature dependence of the upflow velocities to constrain the acceleration and heating mechanism that drives these upflows. We also show that the temporal characteristics of these upflows suggest an episodic driver that sometimes leads to quasi-periodic signals. We suggest that at least some of the quasi-periodicities observed with coronal imagers and spectrographs that have previously been interpreted as propagating magnetoacoustic waves, may instead be caused by these upflows.
\end{abstract}


\section{Introduction}
One of the most overlooked but important problems in solar and stellar physics is the nature of the transport mechanism that is required to supply enough mass from the cool chromosphere to the hot corona to counter the effects of the slow, but steady cooling and draining of the corona on timescales of order half an hour \citep{Schrijver2000}. Ever since their discovery in the late 1800s when ``vertical flames'' where first seen at the limb during total solar eclipse \citep[][]{Secchi1877}, spicules or dynamic jet-like extrusions \citep[][]{Roberts1945,Beckers1968} have received much attention as agents in the transport of energy and mass in the lower solar atmosphere \citep[e.g.,][]{Thomas1948, Pneuman1978, Athay1982}. This is because the mass they carry upward is estimated to be two orders of magnitude larger than that needed to supply mass to the corona and solar wind \citep[e.g.,][]{Athay1982}. One shortcoming was that a signature of these ejecta could not be observed at ``typical'' coronal temperatures \citep[e.g.,][]{Withbroe1983} and so the concept of spicules as a hot mass transport mechanism for the outer solar atmosphere has been largely mothballed.

The Solar Optical Telescope \citep[SOT;][]{Tsuneta2008} of \hinode{} \citep{Hinode} has revolutionized our view of the dynamic chromosphere revealing the existence of at least two types of spicules \citep[][]{DePontieu2007b}. ``Type-I'' spicules are long-lived (3-5 minutes) and exhibit longitudinal motions of the order of 20km/s that are driven by shocks resulting from the leakage of p-modes and convective motions, as well as magnetic energy release in regions around strong magnetic flux concentrations \citep[][]{DePontieu2004,Hansteen2006, DePontieu2007a,Martinez-Sykora2009}. Their contribution to TR emission may be limited to a hot sheath with little of the chromospheric plasma heated to TR temperatures \citep{deWijn2006,Judge2008}. ``Type-II'' spicules, on the other hand, inhabit a truly different dynamic regime than the ``classical'' spicules studied by \citet[][]{Roberts1945} (or others) and have a direct impact on the TR.  These recently discovered spicules show much shorter lifetimes (50-100s) and higher velocities (50-150~km/s), and are likely caused by reconnection \citep[][]{DePontieu2007b}.  Type II spicules often exhibit only upward motion, with the feature fading along its entire length in a matter of seconds, which is suggestive of rapid heating to (at least) TR temperatures.

\citet[][]{DePontieu2009} (hereafter D2009) recently studied active region plage and discovered a direct connection between Type-II spicules and a persistent, faint, but strongly blue-shifted component of emission in hot coronal lines at the magnetic footpoints of loops.  Their work suggests that coronal heating often occurs at chromospheric heights in association with jets that are driven from below and supply the corona with a significant fraction of the required mass flux. 

Here we show that a similar process occurs in the quiet Sun for a wide range of temperatures that cover both the lower and upper TR (or low corona). In contrast to D2009, the current paper is focused on the quiet Sun. We study several aspects of this mass-loading process that were not addressed in D2009. In \S\ 2 we discuss the quiet Sun observations and present a novel analysis technique for UV emission lines observed by SUMER \citep[][]{Wilhelm1995} on \soho{} \citep[][]{Fleck1995}. We perform a detailed study of the strengths and limitations of the line asymmetry analysis. We investigate the effects of blends on the analysis, and compare data observed in both first and second order to show that our conclusions are not impacted by blends. We describe the temporal behavior of the blueward asymmetries and compare them to those of type~II spicules. Most importantly, we put forth a new interpretation of the nature of the emission from the quiet Sun transition region in terms of the mass cycling associated with the heating and cooling of the plasma associated with the type-II spicules. We discuss this new interpration in \S\ 3.

\section{Observations \& Analysis}\label{obs}

In the following we will present line asymmetry analysis of SUMER raster scans for spectral lines of the following ions: \ion{Si}{4}~1402 \AA{}, \ion{C}{4}~1548 \AA{}, \ion{N}{5}~1238 \AA{}, \ion{O}{6}~1031\AA{}, and \ion{Ne}{8}~770 \AA{}, which are formed over a wide range of temperatures (under equilibrium conditions), respectively for $log T= 4.8, 5.0, 5.3, 5.5, 5.8$ \citep[][]{Mazzotta1998}. In addition, we show a sit-and-stare temporal sequence in \ion{N}{5}~1238~\AA{} and analyze the episodic nature of the blue asymmetries.

\subsection{Simultaneous \ion{C}{4} 1548 \AA{} and \ion{Ne}{8} 770 \AA{} observations}
% Fig 2 HERE

The main data set used in the analysis presented is an absolutely calibrated SUMER spectroheliogram in the 1530-1555\AA{} spectral range on September 22 1996 from 00:40-08:28UT for which further details can be found in \citet{Hassler1999} and \citet{Davey2006}. This portion of the solar UV spectrum contains emission lines of singly ionized Silicon (\ion{Si}{2}) with a rest wavelength of 1533.43\AA{}, seven times ionized Neon (\ion{Ne}{8}) at 1540.84\AA{} (in second spectral order) and two lines of three times ionized Carbon (\ion{C}{4}) at 1548.20\AA{} and 1550.77\AA{}. The two \ion{C}{4} lines show identical spectral behavior and so we only consider information from the former as it has better signal-to-noise. For reasons of brevity, low signal-to-noise, and the impact of line blends close to line center, we also exclude the \ion{Si}{2} line from the analysis. This leaves us with two emission lines that span the TR, \ion{C}{4} and \ion{Ne}{8} formed around $10^{5}$ and 6x$10^{5}$K respectively.

\begin{figure}
\plotone{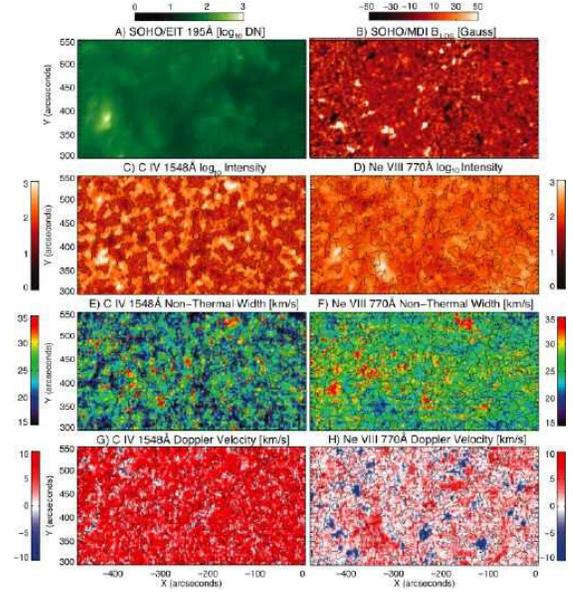}
\caption{Contextual information for this Paper. Top row: the EIT 195\AA{} and composite MDI LOS magnetogram for the SUMER spectroheliogram of September 22 1996. Subsequent rows show the fitted line intensities, non-thermal line widths and absolutely calibrated Doppler velocities of the \ion{C}{4} 1548\AA{} (left) and \ion{Ne}{8} 770.4\AA{} (right) emission lines. In each case the fitted spectral information (panels C-H) are overlaid with the 150 count intensity contour of the \ion{C}{4} intensity as a proxy for the supergranular network boundary. \label{f1}}
\end{figure}

\begin{figure}
%\epsscale{0.7}
\plottwo{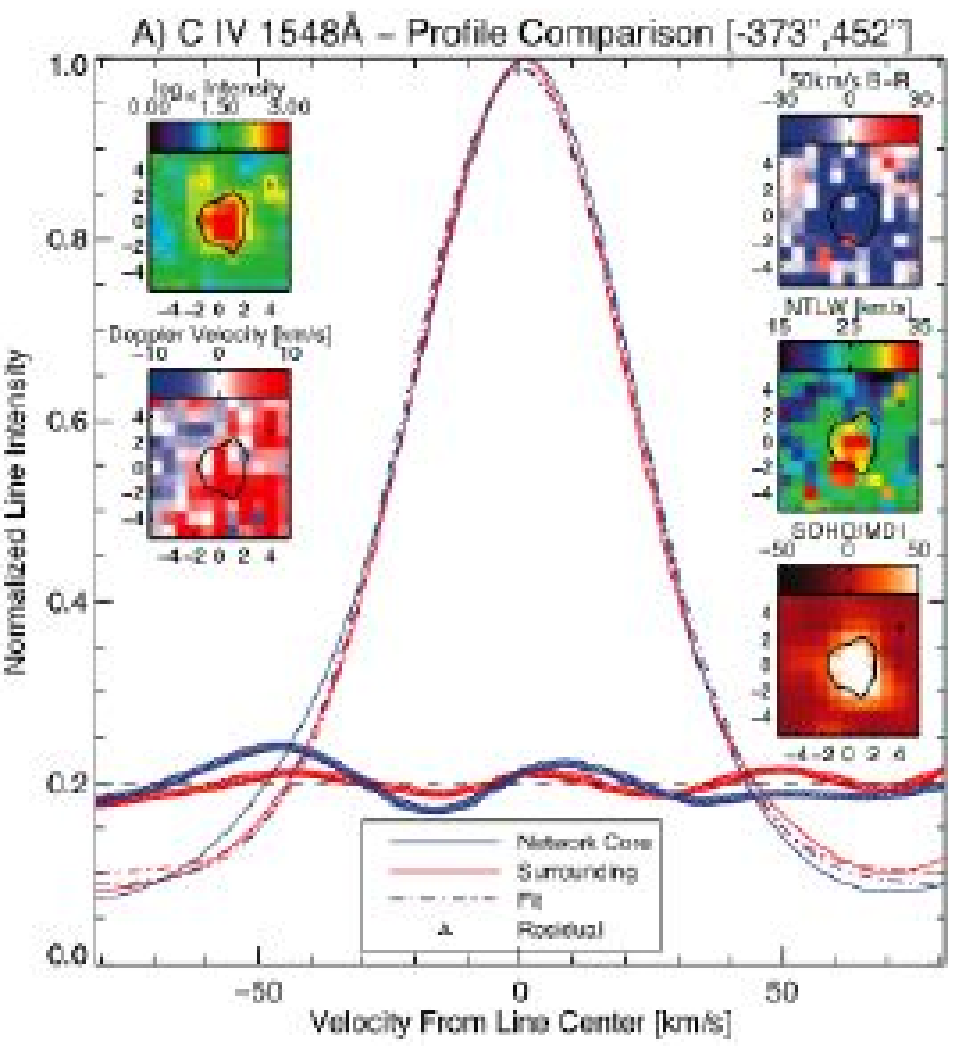}{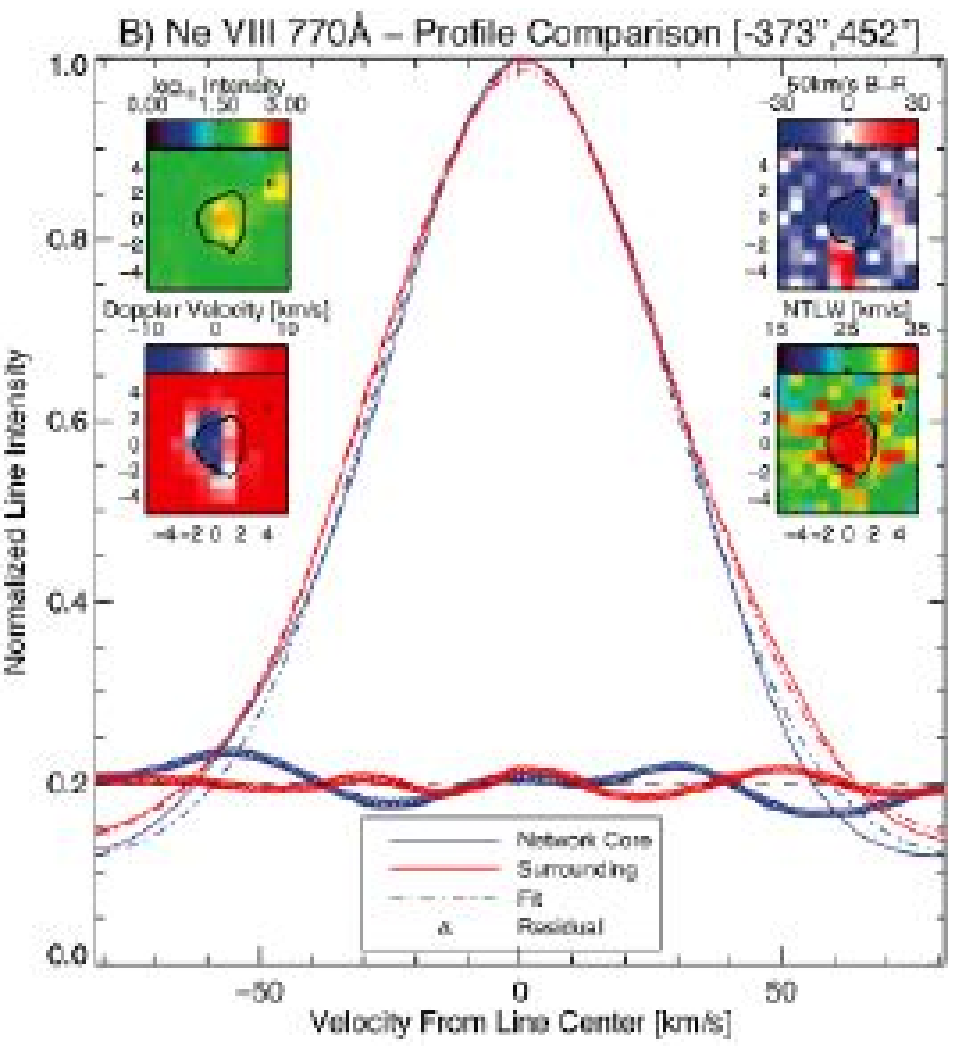}
\caption{Reference network line profiles in \ion{C}{4} 1548\AA{} (left) and \ion{Ne}{8} 770.4\AA{} (right) for the network element centered on x=-373\arcsec, y=452\arcsec. Shown inset in each panel are the intensities, Doppler velocity, profile asymmetry at 50km/s ($\pm$12km/s; see text), $v_{nt}$ and MDI LOS magnetic field strength (in panel A only). The normalized mean profiles are shown for the network (blue solid line), the surrounding inter-network region (red solid line) while the Gaussian fits to those profiles and residuals (profile - fit) are shown as dot-dashed lines and triangular symbols of the appropriate color. The horizontal dashed line is the zero-level of the profile residual [placed at a normalized intensity of 0.2]. \label{f2}}
\end{figure}

For context, Figure~\pref{f1} provides an EIT \citep[][]{Boudine1995} 195\AA{} image (taken at 01:05UT) and a composite MDI \citep[][]{Scherrer1995} line-of-sight (LOS) magnetogram\footnote{The magnetogram is a composite of the four 96-minute cadence full-disk magnetograms taken over the duration of the SUMER spectroheliogram (01:39; 03:15; 04:48; 06:27UT). The composite is needed to partially take into account the considerable evolution of the photospheric magnetic field during the 7 hours of SUMER observations.} as well as the line intensities, non-thermal line widths \citep[$v_{nt}$; determined from the recipe in][]{McIntosh2008} and absolutely calibrated Doppler velocities of the \ion{C}{4} 1548\AA{} and \ion{Ne}{8} 770.4\AA{} lines in the left and right columns respectively. The visibility of the supergranular network differs markedly between \ion{C}{4} (all network contours in Fig.~\pref{f1} are based on \ion{C}{4} intensity) and \ion{Ne}{8} (panel D) in which the network pattern is not obvious. We also see a correspondence between increased values of $v_{nt}$ and the LOS field strength\footnote{We note that there are many locations of enhanced $v_{nt}$ that straddle regions where the magnetic field is expected to be significantly inclined, a topic covered by \citet[][]{McIntosh2009}, or the locations of ``explosive events'', see, e.g., \citet{Innes1997} and \S\ 2.1.}. In the bottom row of Fig.~\pref{f1} the striking prevalence of red-shifted material is evident in the \ion{C}{4} emission \citep[see, e.g.,][and many others]{Gebbie1981} which is enhanced in the supergranular network vertices. In contrast, the \ion{Ne}{8} emission in those same locations show modest blue-shifts or outflows (see Fig.~\pref{f2}), that are typical for quiet Sun \citep{McIntosh2007}.

We employ the technique discussed in D2009 to quantify the asymmetry of the SUMER line profiles. We first perform a single Gaussian fit (with a constant background) to the line profile \citep[][]{McIntosh2008} to determine its centroid position. Then we interpolate the line profile to a wavelength scale that is ten times finer and calculate the total number of counts in the interpolated profile in two narrow range of wavelengths/velocities (two original SUMER spectral pixels wide, or 24 km/s at 1548 \AA) that are symmetrically positioned from the line centroid position, one in the blue and one in the red wing. The difference of the counts in the two wing positions, red-blue (R-B), provides a measure of the asymmetry of the profile at the chosen wavelength or velocity offset. Of course, a measure of zero indicates that the profile is symmetric in that velocity range. The R-B measure is related to the third (skewness) and fourth (kurtosis) moments of the the line profile. For relatively noisy line profiles, such as those studied, the difference of the two wings is a measure of asymmetry, provided that no spectral blends are present in the lines. 

We note that the \ion{Ne}{8} 770.4\AA{} line, seen in second spectral order by SUMER, has two blends of \ion{Si}{1}. As determined from the HRTS spectral atlas \citep[][]{Brekke1993} these blends have equal magnitude, are $\sim$20 times weaker, and are symmetrically distributed at $\sim$20km/s either side of the \ion{Ne}{8} rest position in the quiet sun. Based on the symmetry of the blend positions, their equal intensity and considerably lower amplitude (than \ion{Ne}{8} 770.4\AA{}) we deduce that their net effect on the \ion{Ne}{8} line R-B diagnostic in the quiet sun is negligible\footnote{In general, the presence of a strong blend close to the line core can impact the R-B asymmetry analysis because of its effect on line centroiding. For example, a strong blend in the red wing, close to the core of the line, will shift the line centroid slightly to the red, which can lead to a generalized weak blue asymmetry. Since the \ion{Si}{1} blends are symmetrical, this is unlikely to be the dominant cause of the blueward asymmetries we find in \ion{Ne}{8} 770 \AA{} as observed in second order.}. In a coronal hole the relationship between the three spectral lines is more complex and is discussed at length in a forthcoming Paper \citep[][]{McIntosh2009}. 

%Fig 3 Here

Figure~\pref{f2} illustrates normalized line profiles for a magnetic network element (red solid line) and the surrounding quiet region (blue solid line) centered on coordinates (-373\arcsec, 452\arcsec). We see that for both of the emission lines in the network element there is an excess of emission in the blue wing (similar to what is shown in D2009). The peak of the excess emission is at $\sim$50km/s for \ion{C}{4} and $\sim$55km/s for \ion{Ne}{8}. The surrounding plasma shows little, if any, excess emission. This is clear from the plot of residual emission (triangular symbols that are raised by 0.2 from the lower portion of the plot) between the mean region profile and a single Gaussian profile fit (appropriately colored dot-dashed line). In the network, the amplitude of the asymmetry is of the order of 3-8\% of the peak brightness of the \ion{C}{4} line and 3-5\% of the \ion{Ne}{8}. Note that the excess emission in the blue wing occurs in a region in which the dominant core of the \ion{C}{4} spectral line predominantly shows modest redshifts of order a few km/s. 

\begin{figure}
%\epsscale{0.70}
\plotone{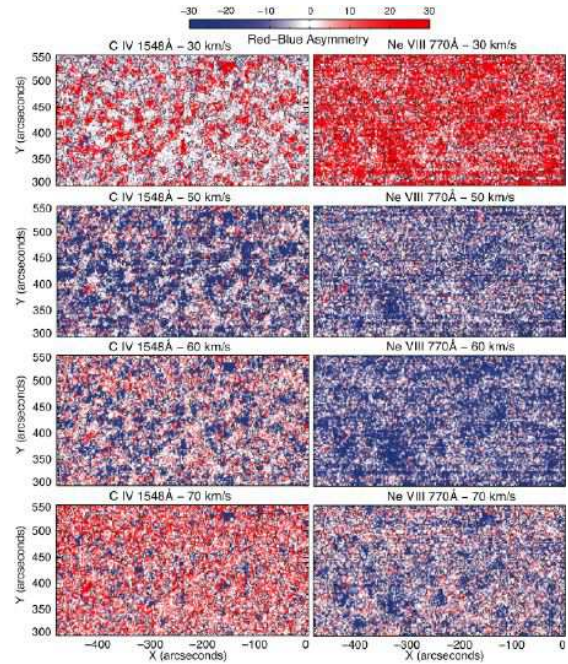}
\caption{Red-blue (R-B) profile asymmetries (in arbitrary units of intensity) for the \ion{C}{4} 1548\AA{} (left) and \ion{Ne}{8} 770.4\AA{} (right) emission lines. Again, for reference, the panels of the figure are overlaid with the 150 count intensity contour of the \ion{C}{4} intensity as a proxy for the supergranular network boundary. The electronic edition of the journal has movies showing the panels of this figure along with those at intermediate velocities and the components of Fig.~\pref{f1}. \label{f3}}
\end{figure}

\begin{figure}
%\epsscale{0.70}
\plotone{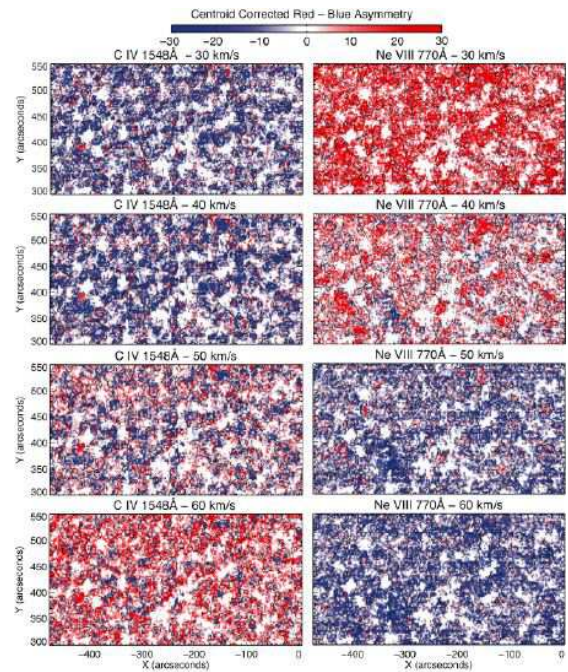}
\caption{Red-blue (R-B) profile asymmetries (in arbitrary units of intensity) corrected for the line centroid positions using the data presented in panels F \& G of Fig.~\pref{f1} respectively for the \ion{C}{4} 1548\AA{} (left) and \ion{Ne}{8} 770.4\AA{} (right) emission lines. For reference the panels of the figure are overlaid with the 150 count intensity contour of the \ion{C}{4} intensity as a proxy for the supergranular network boundary. \label{f4}}
\end{figure}

Calculating the R-B asymmetry measure for the two emission lines over a range of reference velocities from the line centroid, we produce the series of images shown in Fig.~\pref{f3} (and accompanying movies). In both lines at low velocities (below sound speed expected for the relevant formation temperatures) we see that the R-B asymmetry is dominated by the red wing. This is particularly the case in the supergranular boundaries with the asymmetry becoming increasingly large over the strongest supergranular vertex flux concentrations. At higher reference velocities we see that the R-B asymmetry in the supergranular network switches sign indicating that the blue side of the line profile is becoming dominant.  The change in sign is abrupt for \ion{C}{4}, and occurs at about 10-20km/s higher velocities for \ion{Ne}{8}: we see a striking similarity between the two panels when we compare the R-B asymmetry for \ion{C}{4} at 50km/s and \ion{Ne}{8} at 70km/s. 

We should note that the preponderance of red signal in the high velocity ($>70$ km/s) panels of the left (\ion{C}{4}) column of the figure is the result of the analysis sampling the \ion{Si}{1} 1548.72\AA{} line that is 0.5\AA{} (i.e., 100 km/s) to the red of the \ion{C}{4} line center. This line is 4-5 times less bright than the \ion{C}{4} line \citep[see, e.g.,][]{Curdt2001} and we see that strong blue asymmetries still persist in many network locations, ie. indicating that the blue wing of the \ion{C}{4} line is still stronger than the narrow \ion{Si}{1} line at the same velocity in the red wing. This blend reduces the visibility of the \ion{C}{4} asymmetry to higher velocities and can explain some, if not all of the velocity offset between the \ion{C}{4} and \ion{Ne}{8} asymmetries. We also note that a very weak \ion{Si}{1} line occurs at about 100 km/s to the red of line center of \ion{Ne}{8} 770 \AA{} (in second order). This line has an intensity of order only 2\% of the \ion{Ne}{8} intensity \citep{Brekke1993}. It likely plays a role in reducing the visibility of \ion{Ne}{8} asymmetry at very high velocities (but since it is weaker than the blend in \ion{C}{4} it has much less effect).

\begin{figure}
%\epsscale{0.70}
\plotone{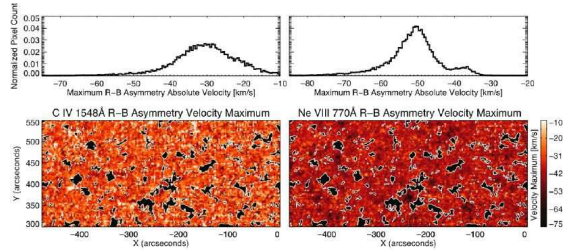}
\caption{Absolute velocities of the blue-wing asymmetry and related histograms in the \ion{C}{4} 1548\AA{} (left) and \ion{Ne}{8} 770.4\AA{} (right) emission lines. The fractional intensities are computed for both emission lines everywhere the \ion{C}{4} intensity is greater than 50 (solid white contours). \label{f5}}
\end{figure}

\begin{figure}
%\epsscale{0.70}
\plotone{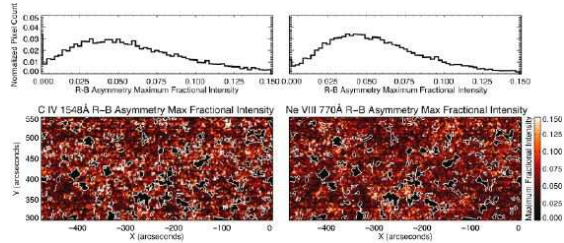}
\caption{Fractional intensities and related histograms for the blue-wing asymmetry in the \ion{C}{4} 1548\AA{} (left) and \ion{Ne}{8} 770.4\AA{} (right) emission lines. The fractional intensities are computed for both emission lines everywhere the \ion{C}{4} intensity is greater than 50 (solid white contours). \label{f6}}
\end{figure}

In order to place the derived R-B asymmetries in some relative context we estimate their maximum amplitude and location in absolute wavelength space. This step requires that we combine the information presented in Figs.~\pref{f1} and~\pref{f3}, using the absolutely calibrated Doppler velocity from the line centroiding (panels \pref{f1}G \& \pref{f1}H) to re-map the R-B asymmetry profiles. At each spatial pixel we adjust the velocity scale of the R-B profile to reflect the measured line centroid velocity; for example, a positive (red) line centroid Doppler shift will shift the entire R-B intensity profile to the red by the corresponding velocity and conversely for a blue Doppler shift of the line centroid. After shifting the R-B profiles we resample the panels of Fig.~\pref{f3} to show the magnitude of the asymmetry in absolute velocity space, these are presented in the panels of Fig.~\pref{f4}. In performing this analysis we can also extract the maximum fractional (relative to the peak line intensity) magnitude of the line asymmetry and its location in absolute velocity space. The plots of these quantities, and their corresponding histograms, can be found in Fig.~\pref{f5} and~Fig.~\pref{f6} respectively where we see that the maximum fractional intensity of the asymmetry in the \ion{C}{4} and  \ion{Ne}{8} lines are of the order of 3-4\% (increasing into the cell interior as the intensity falls off significantly) with velocities that are offset by $\sim$20km/s.

\subsection{\ion{Ne}{8} 770\AA{} first order observations}

To further illustrate the ubiquitous presence of these blue-wing profile asymmetries in the quiet Sun magnetic network we present data that does not suffer significantly from blending with other lines. We also discuss the potential impact of low signal-to-noise on the detection of the subtle asymmetries we have found. 

\begin{figure}
\epsscale{1.10}
\plotone{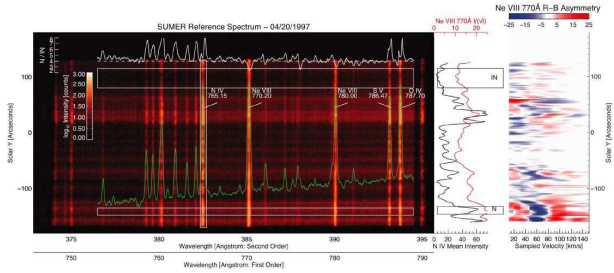}
\caption{The SUMER quiet Sun spectral atlas observation of April 20 1997 00:34UT. The main panel of the figure shows several key measures: the SUMER reduced detector image displaying the spatial variation of several strong UV emission lines, the slit averaged spectrum (green), a vertical box enclosing the \ion{N}{4} 765\AA{} line. Isolating this line, we can compute the averaged TR brightness across the SUMER slit and identify sample regions of (supergranular) network and inter-network (horizontal boxes) labeled ``N'' and ``IN'' respectively. The red curve shows the \ion{Ne}{8} 770\AA{} $I/\sqrt{I}$ measure as an estimate of the spectral quality across the SUMER slit. The white trace along the top of the detector image shows the ratio of the network to inter-network spatially averaged spectra over the spectral domain. The far right panel of the image shows the R-B analysis of the first order \ion{Ne}{8} 770\AA{} emission line along the SUMER slit. \label{f7}}
\end{figure}

\begin{figure}
\epsscale{1.10}
\plotone{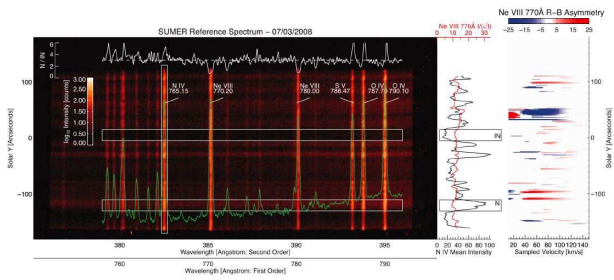}
\caption{The SUMER quiet Sun spectral atlas observation of July 3 2008 12:11UT. The measures derived from the detector image are as for Fig.~\pref{f7}. \label{f8}}
\end{figure}

We consider quiet Sun spectral atlas observations of the \ion{Ne}{8} 770\AA{} line in first spectral order performed by SUMER close to the start of the mission \citep[April 20 1997 00:34UT;][Fig.~\pref{f7}]{Curdt2001} and in a more recent observation \citep[July 3 2008 12:11UT;][Fig.~\pref{f8}]{Tian2009}. These data are taken with extremely long exposures (300s) of SUMER detector B and provide information about the \ion{Ne}{8} 770\AA{} line in first order. This line may be susceptible to \ion{Si}{1} blends in the second spectral order, but is ``clean'' in the first order. We note that the nearest lines in the SUMER spectral atlas are the 769.38\AA{} \ion{Mg}{8} and 771.90\AA{} \ion{N}{3} lines which have $\sim$2$\%$ of the brightness of \ion{Ne}{8} 770\AA{} line in the quiet Sun, and are at 390 and 590 km/s to the blue and red respectively \citep[][]{Curdt2001}. 

Comparing Figs.~\pref{f7} and~\pref{f8} we are immediately drawn to the significant change in the strength, and ``quality'' of the SUMER spectra in the eleven years between the two observations presented. The ratio of network-internetwork intensity (N/IN) across the entire spectral range is reduced by a factor of two, the N/IN profile is very noisy and of low spectral contrast in the later observation, and the peak intensity of many of the emission lines (including \ion{Ne}{8} 770\AA) in this spectral range are decreased by up to a factor of four. Indeed, comparing the cross-slit variation of the \ion{Ne}{8} 770\AA{} spectral quality measure \citep[$I/\sqrt{I}$; a good estimate of the signal-to-noise ratio \-- S/N, see, e.g.][]{Chae1998}, we see a mean value of 11.8, with very little variance (1.5) and very little contrast between the network and inter-network regions in the July 2008 spectra. Compare these values to the April 1997 case where the mean is $\sim$15, with a variance of 4 and regions of the slit with values in excess of 20. Because the strongly blue-shifted component we are interested in is very faint ($\sim 5$\% of the line core), it is clear that the spectral quality will significantly affect our ability to measure/quantify this component. Our analysis suggests that in the later phases of the SOHO mission SUMER spectra with a S/N of the order of 20 (needed to make a robust detection of these faint emission components in the blue wing) are rare even in long exposure observations. This can explain the lack of blueward asymmetries in the first order \ion{Ne}{8} 770 \AA{} spectral profiles (using SUMER) as analyzed recently by \citet[][]{Tian2009}. 

This issue of data quality is illustrated by the results of an R-B analysis of the SUMER \ion{Ne}{8} 770\AA{} spectra in the reference spectra of Figs.~\pref{f7} and \pref{f8}. A complicating factor in this region of the UV spectrum is the linear behavior of the continuum background: even a small gradient can add additional weight to the red wing of the line unless it is taken into consideration when fitting the line center and when assessing the asymmetry of the profile. Therefore, we carefully fit a linear background and perform the R-B analysis on the background-subtracted spectra. The results are shown in the far right panels of Figs.~\pref{f7} and~\pref{f8}. In the April 1997 reference spectrum we see a repetition of the signature observed in Figs.~\pref{f3}, \pref{f4}, \pref{f5} and \pref{f6} (which were for the same spectral line, but in second spectral order): a red wing excess in the network regions at low velocities transitioning to a blue wing excess at higher velocities. The gradual return to red wing excess at very high velocities is likely the result of the linear background which, when not fully removed far from the line core, will lead to a preference for red asymmetries. The fact that we observe blue asymmetries for velocities between 50 and 90 km/s despite this background issue underscores the strength of the blue wing excess\footnote{We note that the coronal hole reference spectrum of \ion{Ne}{8} 770 \AA\ in first order \citep[presented as the blue curve in Fig.~4, pg. 602, of ][]{Curdt2001} likely shows the blue-wing asymmetry as well. We suspect that the ``knees'' visible in the blue wings of the network/internetwork ratio in bright, spectrally isolated, TR emission lines in the spectral atlas data \citep[the green curve in][]{Curdt2001} may be the signature of the network blue profile asymmetry. Further analysis and modeling are required to ascertain the validity and uniqueness of this point.}. In contrast, the R-B signature in the July 2008 spectra is highly dubious: there appears to be little correlation between the patches of signal and any physical structure along the slit. We deduce that this erratic signal is the result of the considerably higher noise level of the spectra taken later in the mission. This degradation of the spectra is a subject that merits further investigation. Reference spectra like those shown have been taken periodically throughout the SOHO mission, so such an analysis is possible, but beyond the scope of this Paper.

\subsection{Raster scans of \ion{Si}{4} 1402\AA{}, \ion{N}{5} 1238\AA{} and \ion{O}{6} 1032\AA{}}

The ubiquity and predominance in network regions of blueward asymmetries for velocities of order 40-100 km/s in the simultaneous \ion{C}{4} 1548 \AA{} and \ion{Ne}{8} 770 \AA{} (in second order) raster scans and the reference spectra in \ion{Ne}{8} 770 \AA{} (in first order) are confirmed by several (non-simultaneous) SUMER raster scans (taken early during the SOHO mission) for three different lines that have high S/N, do not suffer from blends and are formed (under equilibrium conditions) at similar temperatures as \ion{C}{4} and \ion{Ne}{8}. These lines are \ion{Si}{4} 1402.8 \AA{}, \ion{N}{5} 1238.8 \AA{} and \ion{O}{6} 1031.9 \AA{}, formed for $log T = 4.8, 5.3, 5.5$ respectively.

\begin{figure}
%\epsscale{0.55}
\plotone{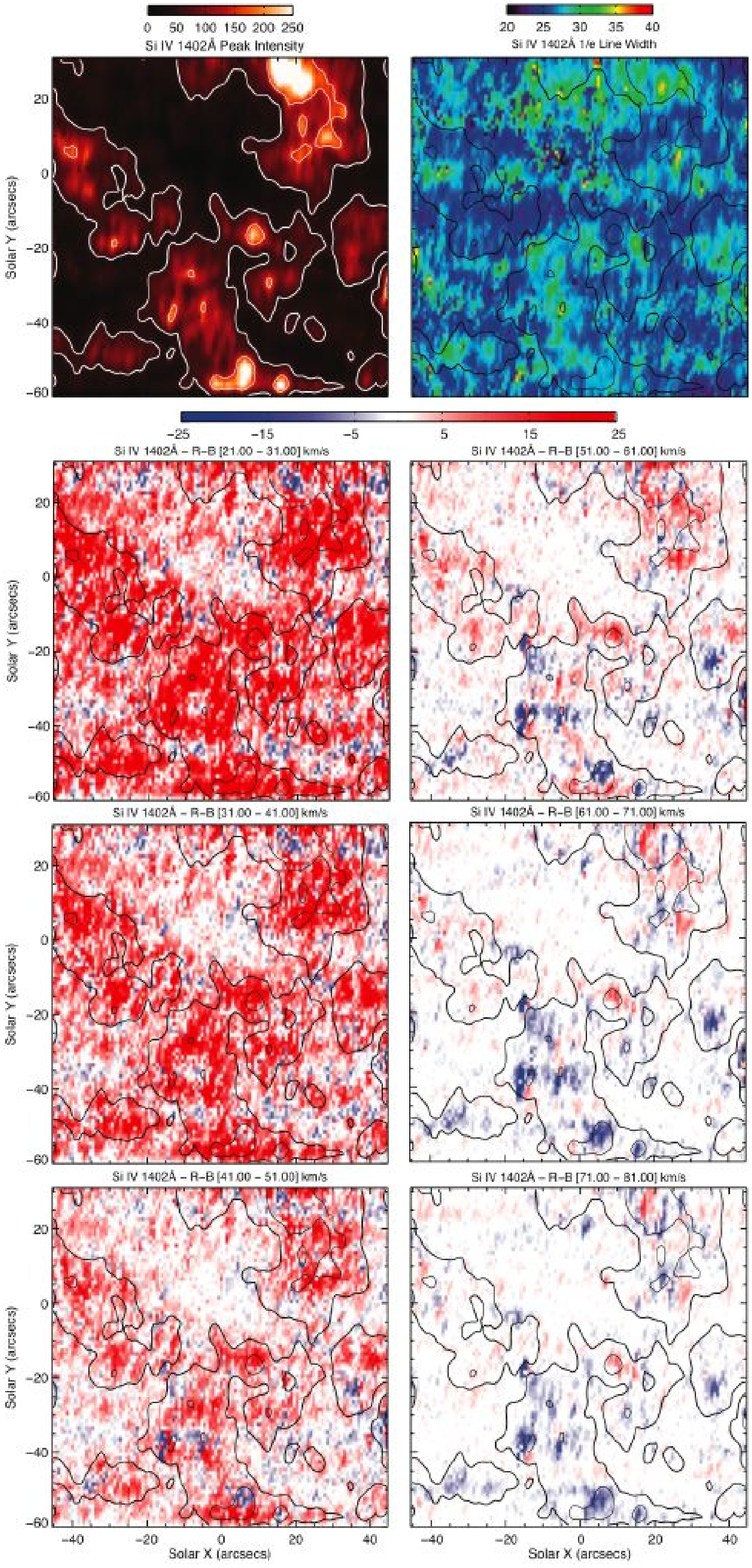}
\caption{The August 6 1996 SUMER quiet Sun, disk center, spectroheliogram in \ion{Si}{4} 1402\AA{} (02:19\--04:18UT \-- 60s exposure). The top row of panels shows the peak line intensity and $1/e$ line width while the R-B panels demonstrate the profile asymmetry in steps from 21 to 81 km/s (left to right, top to bottom). In each case the quantities are overlaid with intensity contours at peak intensity levels of 50 and 150 counts to delineate the supergranular network boundary (thin solid lines) and the strongest network locations (thick solid lines) respectively. The electronic edition of the journal has movies showing the panels of this figure.\label{f9}}
\end{figure}

\begin{figure}
%\epsscale{0.55}
\plotone{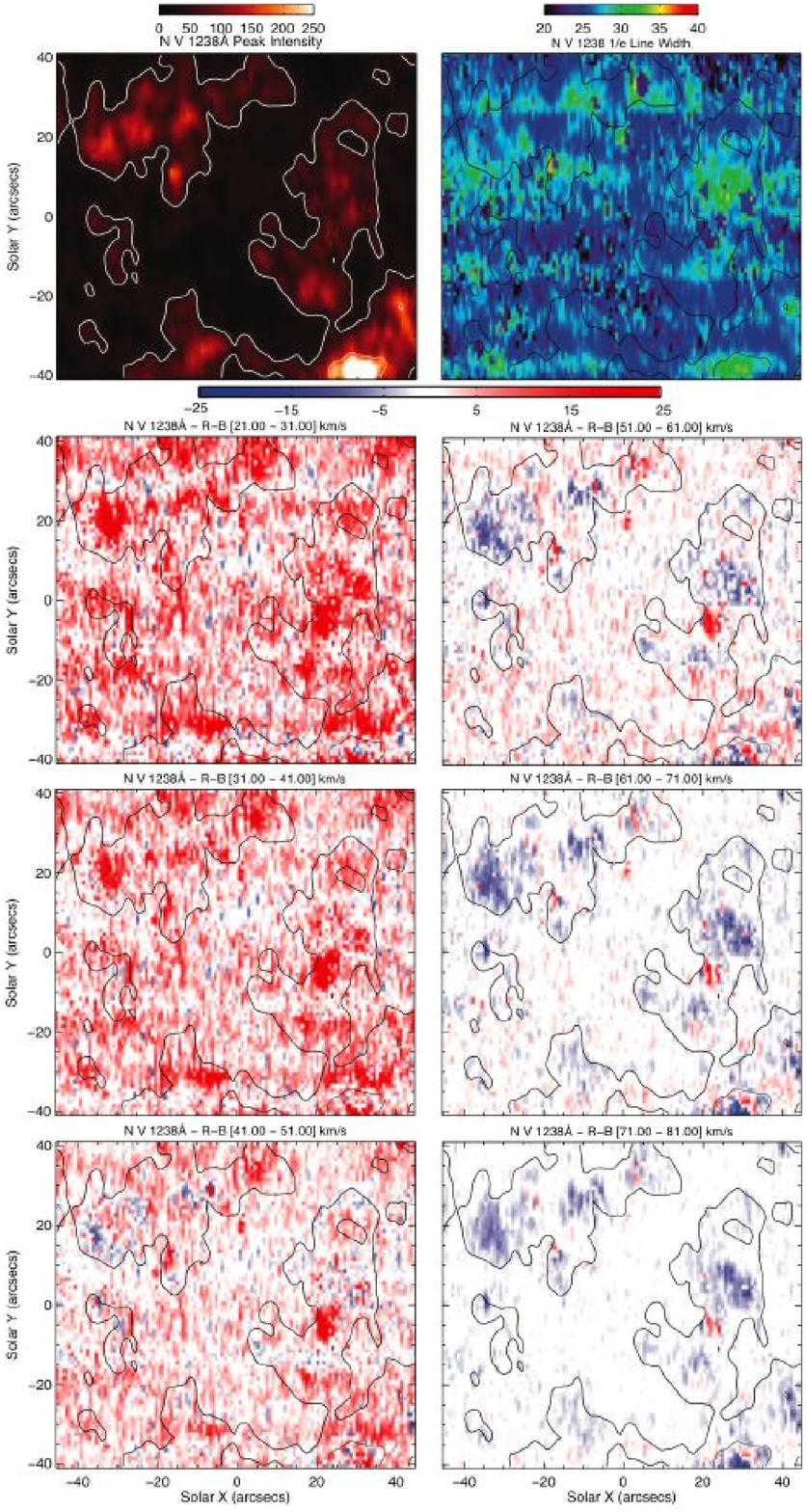}
\caption{The August 6 1996 SUMER quiet Sun, disk center, spectroheliogram in \ion{N}{5} 1238\AA{} (14:49\--15:19UT \-- 10s exposure). The top row of panels shows the peak line intensity and $1/e$ line width while the R-B panels demonstrate the profile asymmetry in steps from 21 to 81 km/s (left to right, top to bottom). In each case the quantities are overlaid with intensity contours at peak intensity levels of 25 and 100 counts to delineate the supergranular network boundary (thin solid lines) and the strongest network locations (thick solid lines) respectively. The electronic edition of the journal has movies showing the panels of this figure. \label{f10}}
\end{figure}

Figure~\pref{f9} shows a raster scan with long exposures (60s) of the \ion{Si}{4} 1402\AA{} line (formed at $T\sim 65,000$ K) taken at quiet Sun at disk center on August 6, 1996. This line does not have any blends in the velocity range shown \citep[according to the SUMER atlas][]{Curdt2001}. Similar to the \ion{C}{4} and \ion{Ne}{8} spectra, the R-B analysis reveals that the network regions (shown with contours based on core intensity of the spectral line) are dominated by red asymmetries for low velocities, with a sudden shift to blue asymmetries from 50 km/s onward. We see significant blue asymmetries even at 80 km/s. 

A very similar picture appears in a disk center raster scan (Fig.~\pref{f10}) taken about 12 hours later, but now in the \ion{N}{5} 1238 \AA{} line \citep[formed at slightly higher temperatures under equilibrium conditions: $\sim$2x$10^{5}$K, ][]{Mazzotta1998}. In the network regions, we see a similar shift from redward asymmetries for velocities less than 50 km/s, to blueward asymmetries for velocities above 50 km/s, up to 80 km/s (the limit of the spectral range for this raster). This line has no known blends in this velocity range. Again, many locations in the network show significant blue asymmetries even at 80 km/s.

\begin{figure}
\epsscale{1.10}
\plotone{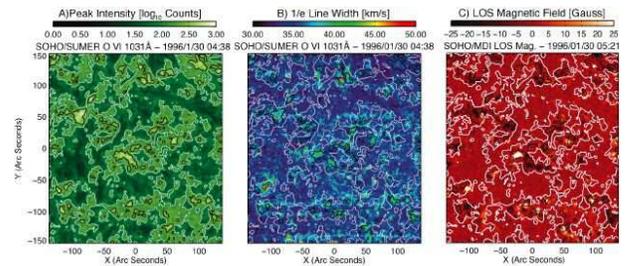}
\caption{Contextual information for the January 30 1996 1031.93\AA{} SUMER spectroheliogram of quiet sun at disk center. From left to right we show the peak line intensity, the $1/e$ of the line profile and an SOHO/MDI magnetogram taken shortly after the end of the SUMER observation (05:21UT). In each case the quantities are overlaid with intensity contours at peak intensity levels of 100 and 250 counts to delineate the supergranular network boundary (thin solid lines) and the strongest network locations (thick solid lines) respectively. \label{f11}}
\end{figure}

\begin{figure}
%\epsscale{1.00}
\plotone{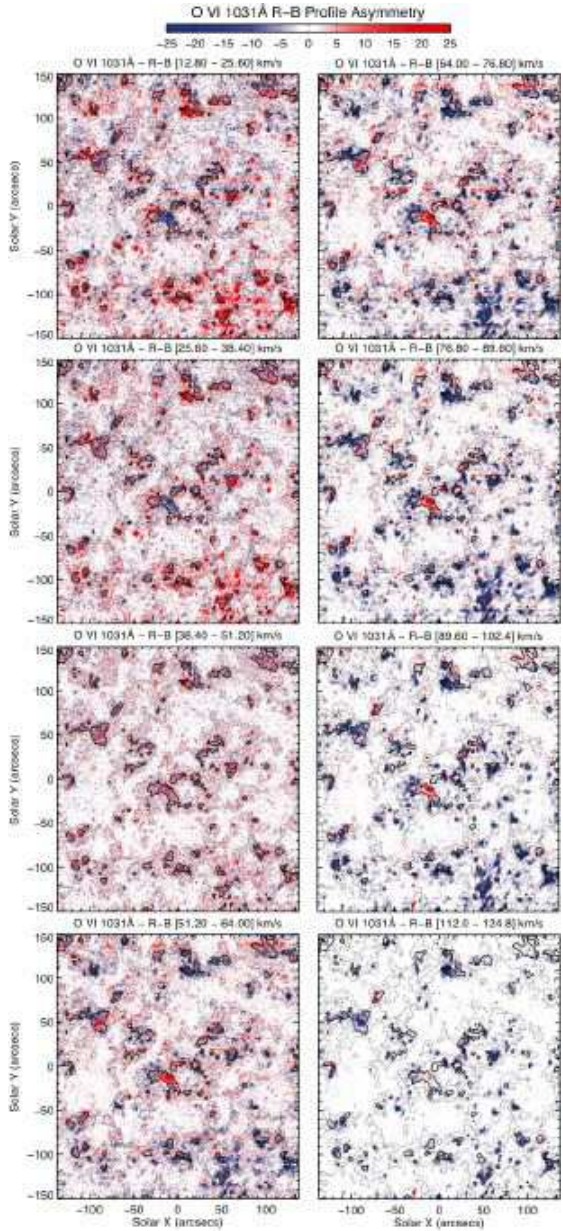}
\caption{Red-blue (R-B) profile asymmetries (in arbitrary units of intensity) for the \ion{O}{6} 1031.93\AA{} emission line. The panels of the figure are overlaid with intensity contours at peak intensity levels of 100 and 250 counts to delineate the supergranular network boundary (thin solid lines) and the strongest network locations (thick solid lines). The electronic edition of the journal has a movie showing the panels of this figure along with those at intermediate velocities and the components of Fig.~\pref{f11}. \label{f12}}
\end{figure}

And finally, we consider an R-B analysis of the \ion{O}{6} 1031.92\AA{} SUMER observation taken on January 30 1996 from 04:38-04:57UT. According to the SUMER spectral atlas \citep[][]{Curdt2001} the \ion{O}{6} 1031.92\AA{} line \citep[formed in equilibrium at $\sim$4x$10^{5}$K,][]{Mazzotta1998} has one very weak blend ($<$0.5\% of the \ion{O}{6} intensity) in the form of the 1031.39\AA{} \ion{S}{2} line. At this wavelength, any effect of the blend in the blue wing will appear at $\sim$154km/s from the line center, considerably beyond our analysis window and so we consider it to be negligible. Fig.~\pref{f11} presents, from left to right, the peak line intensity, $1/e$-width of the \ion{O}{6} line for comparison with a SOHO/MDI line-of-sight magnetogram taken at 05:21UT, in each case the thick and thin contours delineate the brightest network elements and the (approximate) supergranular network boundary, through 250 and 100 count peak intensity contours derived from panel A. Figure~\pref{f12} shows the results of the R-B analysis of this line, using the steps discussed above. As before, we observe that the strong magnetic regions that compose the supergranular network exhibit a pronounced behavior. These regions favor a red-shifted excess of material at lower velocities that is replaced by a blue-shifted component appearing at about 50km/s with a 5-10\% magnitude relative to the peak intensity that persists, gradually fading. Therefore, the  \ion{O}{6}  line has a qualitatively similar behavior to what we have observed above in \ion{Si}{4}, \ion{C}{4}, \ion{N}{5} and \ion{Ne}{8}. As an aside, we also notice a very high velocity blue component in certain locations (112-125km/s - panel H). Within the limitations of a slowly rastered image and a magnetogram taken 20 minutes later we assess that these locations lie typically (but not exclusively) at the edge of strong magnetic elements, are co-spatial with locations of significant profile broadening ($>50$km/s, see panel B of Fig.~\pref{f11}), and possibly correspond to ``explosive events'' in the periphery of the magnetic network \citep[][]{Innes1997}.

In summary, the raster scans in these three, blend-free, high S/N spectral lines show the same blueward asymmetries for velocities of order 40-100 km/s in network regions as those we found in the \ion{C}{4} and \ion{Ne}{8} spectra. These results strongly suggest that the effect of systemic blends on our results is unlikely to cause all observed asymmetries, and that strong upflows of 40-100 km/s seem to occur all over the quiet Sun in and around network regions, for lines formed (under equilibrium conditions) at temperatures from the low TR to the upper TR and low corona. These upflows are similar in amplitude and nature as those found by D2009 in plage regions, which are associated with chromospheric type-II spicules.

\subsection{Time Series of \ion{N}{5} 1238 \AA{}}

\begin{figure*}
\epsscale{1.}
\plotone{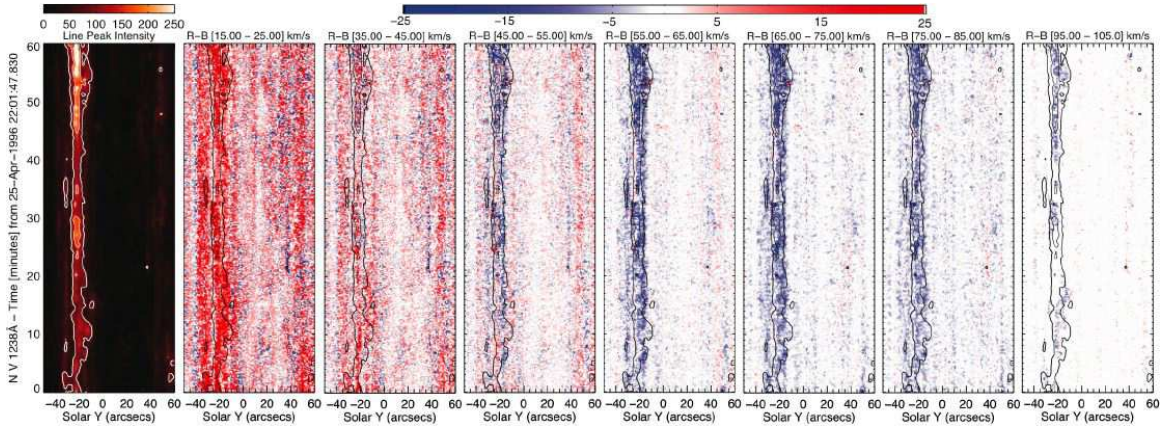}
\caption{The April 25 1996 SUMER quiet Sun, disk center, timeseries in \ion{N}{5} 1238\AA{} (22:01\--23:01UT \-- 7.5s exposure). The panels, from left to right, show the peak line intensity and the R-B profile asymmetry in steps from 15 to 105 km/s. The panels are overlaid with intensity contours at peak intensity levels of 75 and 150 counts to delineate the supergranular network boundary (thin solid lines) and the strongest network locations (thick solid lines) respectively. The electronic edition of the journal has movies showing the panels of this figure. \label{f13}}
\end{figure*}

The raster scans shown in the previous sections clearly show a preference for the blueward asymmetries to occur in and around network regions. However, all of these scans take hours to complete so that temporal characteristics of the upflows cannot be determined. We use a sit-and-stare sequence of a quiet Sun region at disk center taken with SUMER on 25 April 1996 to illustrate the temporal behavior of the upflows. Fig.~\pref{f13} shows a one hour long sequence of \ion{N}{5} 1238 \AA{} spectra that cover a quiet Sun region, including a network region at y=-30\arcsec to y=-20\arcsec. The intensity of the line is slowly changing in the bright network region over the course of the one hour long timeseries. The R-B analysis once again shows the redward asymmetry for velocities less than 50 km/s which shift rapidly to blueward asymmetries for velocities in excess of 50 km/s. Blueward asymmetries dominate from 50 to 100 km/s in the network region. In the internetwork regions very little asymmetry is found for these velocities. The right panels of Fig.~\pref{f13} clearly show that the blueward asymmetries are highly variable in time, even while they continue to occur in and around the network region. This data shows that the upflows of 50-100 km/s are apparently episodic in nature.

\begin{figure}
\epsscale{0.75}
\plotone{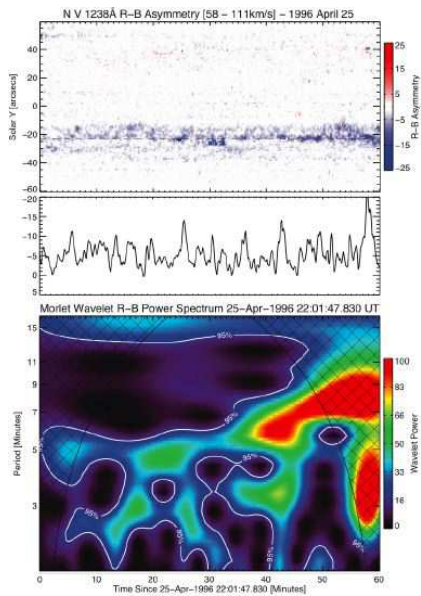}
\caption{Wavelet analysis of the \ion{N}{5} 1238\AA{} R-B (58-111km/s) timeseries (top panel, see e.g, Fig.~\pref{f13}). The R-B timeseries at $y = -22$\arcsec{} are selected for wavelet analysis (center panel) and the resulting wavelet power spectrum is shown (bottom panel). The bottom panel shows the cross-hatched cone of influence and contours of the 90\% significance level in the timeseries. \label{f14}}
\end{figure}

Further analysis of this time series shows that the characteristic timescale on which these upflow events seem to recur is of order 3-15 minutes. This is illustrated in Fig.~\pref{f14} which shows, as a function of time and space, the episodic nature of the R-B asymmetry of \ion{N}{5} 1238 \AA{} (summed from 58 to 111 km/s) for the same timeseries. The middle panel shows the R-B timeseries at $y=-22$\arcsec{} with clear peaks of blueward asymmetry occurring, and typical events lasting of order 1-3 minutes. These events apparently recur on timescales of 3-15 minutes. This is confirmed by a wavelet analysis (using the Morlet wavelet) of the \ion{N}{5} 1238\AA{} R-B (58-111km/s) timeseries: the resulting wavelet power spectrum (bottom panel) shows significant (at 95\% significance level) power throughout the timeseries at periods of order 3-5 minutes, as well as some at 15 min. 

The lifetime of these upflow events agrees surprisingly well with those of type~II spicules \citep[of order 10-100 s][]{DePontieu2007b}. This is further supporting evidence that these quiet Sun events are similar to those seen by D2009 in plage regions (which were directly associated with type~II spicules). It is also very interesting that the upflow events in this particular network region occur quasi-periodically, or at least seem to occur on typical timescales of 3-15 minutes, timescales that are similar to granular (and other photospheric) timescales. The quasi-periodicity of the R-B asymmetries is further evidence that the periodicities observed in coronal imaging and spectral timeseries are not necessarily signs of propagating magneto-acoustic waves in the corona \cite[e.g.][]{deMoortel2002,Wang2009}, but may (sometimes) also be caused by quasi-periodic, high-speed, faint upflows associated with type~II spicules that are heated to coronal temperatures. This topic is further discussed by \citet{McIntosh2009b}.

\section{Discussion \& Interpretation}\label{discuss}
We observe high speed upflows in emission lines that are formed, under equilibrium conditions, in the low, middle, and upper TR (or low corona). The upflows show a similar spatial pattern in all lines, preferentially occur around the network regions, and have velocities of order 40-100 km/s. The emission of the upflowing plasma is faint at $\sim$ 5\% of the peak intensity of the line. It is possible that these upflows are visible as propagating disturbances in EUV \citep[][]{Schrijver1999} and soft X-Ray coronal images \citep[][]{Sakao2008}: the velocity of the measured weak episodic jets are in the same range as the apparent velocities of the disturbances in imaging data. A connection to the chromospheric activity and/or spectroscopic measurements is studied by \citep{McIntosh2009b} for active regions.

What causes these upflows? Many authors have discussed blueward excursions of TR lines in the past \citep[e.g.][]{Dere1989}, with most of those observations being linked to explosive events: violent, often bi-directional jets caused by reconnection \citep{Chae1998}. What we report on here is not an occasional disturbance, like an explosive event, but a much more ubiquitous and fainter phenomenon: at any single moment, most of the network regions show these high speed upflows. We believe that these upflows are related to the blueward asymmetry and line broadening that has been reported before for TR lines in the quiet Sun network \citep[e.g.][]{Peter2000}. The latter suggested, without the benefit of the R-B measure, that hot coronal funnels cause this faint component, with the dominant core of the line emitted in small-scale, low-lying coronal loops. Our interpretation is very different. The quiet Sun upflows are very similar in location, velocity (50-120 km/s), intensity (5\% of line core) and temporal behavior to the faint, coronal upflows observed with Hinode/EIS above active region plage \citep[][D2009]{Hara2008}. The latter have been directly linked to the heating to coronal temperatures of chromospheric type II spicules, which dominate the upper chromosphere in plage and quiet Sun network and develop velocities of order 50-150 km/s. D2009 estimated that the mass flux injected to the corona by these jets is significant for the mass balance of active region coronal loops. We suggest that the upflows we observe here in \ion{Si}{4}, \ion{C}{4}, \ion{N}{5}, \ion{O}{6}, and \ion{Ne}{8} are the quiet Sun equivalent of this process.

%Fig. 5? HERE

\begin{figure}
\plotone{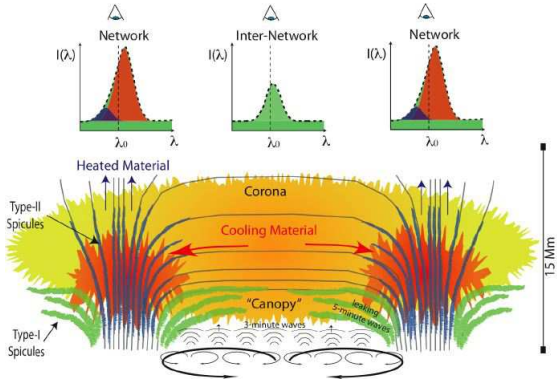}
\caption{Conceptual view of TR emission in network and inter-network.\label{f15}}
\end{figure}

In Fig.~\pref{f15} we provide a cartoon of the TR structure consistent with our observations and the analysis of D2009. What we observe as the transition region is a conveyor belt of sorts, one that includes frequent bursts of rapidly upflowing ionized hot plasma (in the form of thin, dynamic type II spicules), and recycles the material that is slowly cooling out of the corona, losing energy through radiation and producing the well known prevalent TR red-shift, as well as a host of other TR phenomenology discussed above \citep[e.g.,][]{Gebbie1981, deWijn2006, Judge2008}. This figure is very different from Fig.~8 of \citet{Peter2000}. We interpret the supergranular magnetic network as a truly dynamic environment of very high speed ionized plasma upflow (in the form of type II spicules) with a mix of instantaneous and gradual mass return from the corona {\em instead} of a component due to ``hot coronal funnels" rooted in supergranular boundaries. While low-lying, cooler loops \citep[e.g.,][]{Dowdy1986} may well play some role\footnote{\citet{Judge2008a} recently questioned this view of transition region structure.} in the TR emission \citep{Peter2000} our analysis has demonstrated that the role of spicules is fundamental in understanding the observed TR signature. It is possible that the small network core blueshifts observed in \ion{Ne}{8} may be the result of a more gentle evaporative mass loading due to coronal energy deposition \citep[e.g.,][]{Klimchuk2006,Tomczyk2009} in the upper TR. Regardless of the interpretation, our results suggest that approximating the observed spectral line profiles with (even multiple) Gaussians is profoundly limiting, and that other ways to consistently model the observed asymmetric plasma velocity distributions are essential \citep[e.g.,][]{Scudder1992a}. 

Our conceptual view of the TR is related to the model put forth by \citet{Pneuman1978} who suggested that a small fraction of spicules are heated to coronal temperatures and provide the corona with hot plasma. To constrain the spicular mass flux into the quiet Sun corona we follow the approach of D2009, appealing to the conservation of mass and observed emission measures in the system and noting that the latter are similar to those observed in active regions - upflow components with $\sim 5\%$ of the core intensity. D2009 found that the density of the coronal component of the spicules could be expressed as $\rho_i \approx \rho_c /(3\epsilon_T^2)$ where $\rho_c$ the coronal density, and $\epsilon_T$ is the fraction of total coronal spicular density at a temperature of $T$. If we assume that only a fraction $\epsilon_T=0.25-0.5$ of the spicular coronal mass flux is observed in \ion{Ne}{8}, then the only difference with the analysis of D2009 is that the coronal density in quiet Sun is less than that of active regions (say, 3x10$^8$ instead of $10^9$ cm$^{-3}$). As a result the coronal densities $\rho_c$ of hot spicular matter are of order $3 - 15 \times 10^8$ cm$^{-3}$. This implies that the heated spicules can fill the quiet Sun corona even if only 0.3-1.5\% of the chromospheric spicule mass reaches coronal temperatures. D2009 also estimate that the number of spicules per arcsec$^2$ occurring within the network over the lifetime of a spicule is $N_n \approx 2 \epsilon_T^2  \, 725^2/\delta_i^2$, where $\delta_i \approx 250$km the diameter of a typical spicule. Since the assumptions about the coronal scale height, expansion from network to corona, velocity and diameter of spicules are of the same order for active region and quiet Sun, we estimate a density of spicules of order 1-4 per arcsec$^2$ within a network region. Is this number compatible with Hinode/SOT observations of type II spicules at the limb? \citet{Rouppe2009} report that they see at least between 1 and 3 of these spicules per linear arcsec at the limb (at a height of 5,000 km). \citet{Rouppe2009} indicate that when observing at 5,000 km height above the limb, line-of-sight superposition implies that one can see features emanating from network regions over a line-of-sight that is of order 200 arcseconds long. Such a line-of-sight in principle could cross network at the edge of of $\sim 4$ supergranules. However, the network around supergranular cells has a relatively low filling factor of order 0.25, as shown in, e.g., Fig.~12 of \cite{Rouppe2009}. This means that our estimate of 1-4 spicules per arcsec$^2$ within a network region fits in well with the observed number of spicules (1-3 per linear arcsec) that is observed with Hinode at the limb (since the number of supergranule cells multiplier is cancelled by the low network filling factor). We conclude that the observed number of spicules at the limb is fully compatible with a scenario in which the heated component of type II spicules observed in TR emission can play a significant role in the mass balance of the quiet corona.

The impact of these spicules on the coronal energy balance can be estimated by calculating the flux associated with the internal energy of the mass flux calculated in the above. We find that the energy flux is given by $N_i \rho_c c_v T v_i$, with $N_i$ the number of spicules that occur at any location over the spicule lifetime, $\rho_c$ the mass density of the spicule component that is heated to coronal temperatures, $c_v$ the specific heat at constant volume, $v_i$ the upflow velocities in the spicules, and $T$ the temperature the coronal spicular component reaches. D2009 estimated that $N_i \approx 2 \epsilon_T^2$. If we assume that all the plasma in these spicules is heated to 1~MK (of order the formation temperature of Ne VIII, this implies $\epsilon_T \approx 1$), and use $\rho_c = 3 \times 10^8$ cm${-3}$, $v_i \approx 100 $ km s$^{-1}$, we find an energy flux of order $7 \times 10^5$ erg cm$^{-2}$ s$^{-1}$. This is of the same order of magnitude as the estimated conductive and radiative losses \citep[$3 \times 10^5$ erg cm$^{-2}$ s$^{-1}$ for quiet Sun,][]{Priest1982}.
 
Several issues are unresolved in the scenario presented here. The first is the role that spicules play in the mass balance of the lower TR. Observations from Hinode/SOT indicate that many are heated above 20,000K \citep[][]{DePontieu2007b}, but the peak temperature that the bulk of the spicular mass reaches is unclear. \citet{Pneuman1978} suggest that the fraction of spicular mass reaching a certain TR temperature $T$ is a steeply declining function of $T$, consistent with the rapid decrease of differential emission measure from 10,000 to 100,000K. Simultaneous observations with high spectral, temporal and spatial resolution of the chromosphere and TR at high signal-to-noise ($>$20) are necessary to resolve this issue. With the recent selection of the Interface Region Imaging Spectrometer (IRIS) such data should be forthcoming. 

A second issue relates to the formation and heating mechanism of type II spicules themselves. Insight into the magnetic geometry required for the energy release and how the plasma itself is heated and accelerated along the structure of the spicule may help explain why we see a velocity difference in the upward velocity components between the formation conditions of \ion{Si}{4}, \ion{C}{4}, \ion{N}{5}, \ion{O}{6} and \ion{Ne}{8} as well as those between the quiet and active atmosphere (with upflows stronger in the latter -- up to 150 km/s). Certainly, the temporal behavior points to a driver that leads to episodic upflows on characteristic timescales of granulation, something also seen by \citet{Rouppe2009} who report on the discovery of the disk-counterpart of type~II spicules.

Finally, based on the observational evidence and interpretation presented in this paper, efforts to better understand type II spicules may lead to a clearer picture of the impact of non-equilibrium ionization on the upper portion of the solar atmosphere.

\acknowledgements
We appreciate the comments of Alfred de Wijn, Karel Schrijver, the anonymous referee, and the many colleagues who have commented on the presentation of these results at recent meetings. We would like to thank Jeneen Sommers for providing the January 30 1996 SOHO/MDI data. This work is supported by NASA grants NNX08AL22G and NNX08BA99G to SWM and BDP. The National Center for Atmospheric Research is sponsored by the National Science Foundation. The \soho{} is a mission of international cooperation between ESA and NASA. SUMER is supporerted by DLR, CNES, NASA and the ESA PRODEX Program.

\end{document}